BRIEF COMMUNICATION ARISING

GLASS BEHAVIOUR

# Poisson's ratio and liquid's fragility?

Arising from: V. N. Novikov & A. P. Sokolov, *Nature*, **431**, 961 – 963 (2004)

Lack of a reliable theory of glass physics has led to pursuit of correlations between various glass or viscous liquid parameters, and the slope $m$ of log(viscosity) against $T_g/T$ plot extrapolated at the glass transition temperature $T_g$, termed 'fragility', has been used as one of the parameters. Novikov and Sokolov[1] concluded that $m$ of a liquid varies linearly with the ratio of the instantaneous bulk and shear moduli, $K_\infty/G_\infty$, of its glass according to the relation, $m = 29(K_\infty/G_\infty - 0.41)$. Because of the obvious importance of the elastic properties of a glass, here we investigate the basis for this relation[1] and find that its premise is flawed for two reasons, (a) unjustifiable preference for an empirical variation of $m$ with elastic properties and (b) selected use of glasses. When more glasses are considered in the same manner, $m$ does not seem to be linearly related with $K_\infty/G_\infty$.

The square of the ratio of the propagation velocity of longitudinal and transverse ultrasonic (or hypersonic) waves through an isotropic medium, $(v_l/v_t)^2$, yields $K_\infty/G_\infty = [(v_l/v_t)^2 - (4/3)]$. Firstly, we present the data of Fig. 2, Ref. 1, as a plot of $m$ against $K_\infty/G_\infty$ and against $v_l/v_t$ in Fig. 1, panel (A), with the incorrectly plotted data point indicated. (Note that the glasses are binary component, molecular, hydrogen bonded, and network structure type.) Straight lines show least square linear fits to the data. It is evident that the linear relation fits the data equally well whether $m$ is plotted against $K_\infty/G_\infty = [(v_l/v_t)^2 - (4/3)]$, or is plotted against $v_l/v_t$. There seems to be no reason why a linear relation between $m$ and $K_\infty / G_\infty$ had been preferred in Ref. 1 over a linear relation between $m$ and $v_l / v_t$.

Secondly, for a more detailed investigation of the available data on glasses listed in Table I, we present the plot of $m$ against $K_\infty/G_\infty = [(v_l / v_t)^2 - (4/3)]$ and against $v_l / v_t$ in Fig. 1, panel (B), with the 13 data points taken from Ref. 1. It is evident that there is no linear relation between $m$ and the elastic properties of glasses. The glasses are grouped into three types, inorganic, organic and metallic. The data also show that $m$ does not linearly increase with $K_\infty/G_\infty$, even within one group of glasses.

Evidently, both a preference for a linear variation of $m$ with $K_\infty/G_\infty$ (or $v_l^2/v_t^2$ ) over a linear variation of $m$ with $v_l / v_t$, and the limited data used in the correlation have led to an unreliable conclusion in Ref. 1. Therefore, the physical significance of the subsequent discussion of the glass and liquid properties based upon this relation[1] is questionable.

Our findings are significant because Poisson's ratio of a glass is important for use in technology and in structural design. It is a measure of the change in lateral dimensions on elastically loading an object and it is defined as, $\mu_{Poisson} = \frac{1}{2} - \frac{3}{(6K_\infty / G_\infty) + 2}$. The conclusion that "fragility of a liquid is fully determined just by the Poisson ratio of its glass"[1] gives hope to a glass technologist that controlled modification of the composition, that determines a glass melt's $m$, can be used to control its Poisson ratio. Unfortunately the conclusion is untrue.

**Spyros N. Yannopoulos*, G. P. Johari†**

* Foundation for Research and Technology Hellas – Institute of Chemical Engineering and High Temperature Chemical Processes (FORTH / ICE-HT), P.O. Box 1414, GR–26504 Patras, Greece, and † Department of Materials Science and Engineering, McMaster University, Hamilton, Ont. L8S 4L7, Canada

sny@iceht.forth.gr

**Table I:** List of 50 glasses whose data are used here. The first number in brackets is the reference for the *m* value, and the second one is the reference of the elastic properties. In cases, especially for metallic glasses, where several *m* values were reported we used an average value.

| ***Inorganic glasses*** | ***Ref. No*** |
|---|---|
| $GeS_2$ | [2, 3] |
| NBS710 (70.5$SiO_2$ 8.7$Na_2O$ 7.7$K_2O$ 11.6CaO (in wt %) | [4, 5] |
| ZBLA (58$ZrF_4$ 33$BaF_2$ 5$LaF_3$ 4$AlF_3$) | [6, 5] |
| ZBLAN20 (53$ZrF_4$ 20$BaF_2$ 4$LaF_3$ 3$AlF_3$ 20NaF) | [6, 5] |
| HBLAN20 (53$HfF_4$ 20$BaF_2$ 4$LaF_3$ 3$AlF_3$ 20NaF) | [6, 5] |
| 2$BiCl_3$-KCl | [5, 7] |
| $As_2S_3$ | [8, 9] |
| $As_2Se_3$ | [10, 9] |
| $As_xS_{100-x}$ (x=5, 10, 20, 30) | [11, 12] |
| 15LiCl-85$H_2O$ | [13, 14] |
| $Ca(NO_3)_2$-8$H_2O$ | [15, 15] |
| BSC (borosilicate) 70$SiO_2$ 11$B_2O_3$ 9$Na_2O$ 7$K_2O$ 3BaO (wt %) | [16, 17] |
| $CaAl_2Si_2O_8$ (anorthite) | [4, 18] |
| $CaMgSiO_6$ (diopside) | [4, 18] |
| 20$K_2O$-80$B_2O_3$ | [19, 20] |
| 45$B_2O_3$-55$SiO_2$ | [19, 21] |
| 33$Li_2O$-67$SiO_2$ | [19, 22] |
| 33$Na_2O$-67$SiO_2$ | [22, 22] |
| x$Na_2O$–(100-x) 95$GeO_2$, (x=5, 10, 30) | [22, 22] |
| ***Metallic glasses*** | |
| $Pd_{77.5}Si_{16.5}Cu_6$ | [23, 24] |
| $Pt_{60}Ni_{15}P_{25}$ | [23, 24] |
| $Zr_{41.2}Ti_{13.8}Cu_{12.5}Ni_{10}Be_{27.5}$ | [25, 26] |
| $P_{39}Ni_{10}Cu_{30}P_{21}$ | [25, 26] |
| $Ce_{70}Al_{10}Cu_{10}Ni_{10}$ | [27, 27] |
| $Pd_{40}Ni_{40}P_{20}$ | [25, 28] |
| $Zr_{46.75}Ti_{8.25}Cu_{7.5}Ni_{10}Be_{27.5}$ | [25, 29] |
| $Pd_{64}Ni_{16}P_{20}$ | [25, 30] |
| $Pd_{16}Ni_{64}P_{20}$ | [25, 30] |
| $Zr_{50}Cu_{50}$ | [31, 32] |
| $Pr_{60}Cu_{20}Ni_{10}Al_{10}$ | [33, 33] |
| ***Alkali-Borate glases*** | |
| x$Li_2O$–(100-x)$B_2O_3$ (x=10, 20, 30) | [34, 20] |
| x$Na_2O$–(100-x)$B_2O_3$ (x=10, 15, 20, 25, 30, 35, 40) | [34, 20] |
| ***Bismuth-Borate Glasses*** | |
| x$Bi_2O_3$–(100-x)$B_2O_3$ (x=20, 25, 30, 35, 45) | [22, 35] |

## Figure Caption

**Figure 1. Relation between *m* and elastic properties of glasses.** Plots of *m* showing a preference for data fitting in panel (A), and lack of linear relation with *m* in panel (B). The data in panel (A) are from Ref. 1, showing the erroneously plotted red dot for glycerol and blue dot for m-toluidine. In panel (B), the black symbols are from Ref. 1, and coloured symbols are data for glasses listed in Table I. Solid circles refer to various inorganic glasses; open triangles to alkali borates; open diamonds to bismuth borates; closed diamonds, to alkali-germanates; half-filled circles to arsenic selenides; and open circles to metallic glasses.

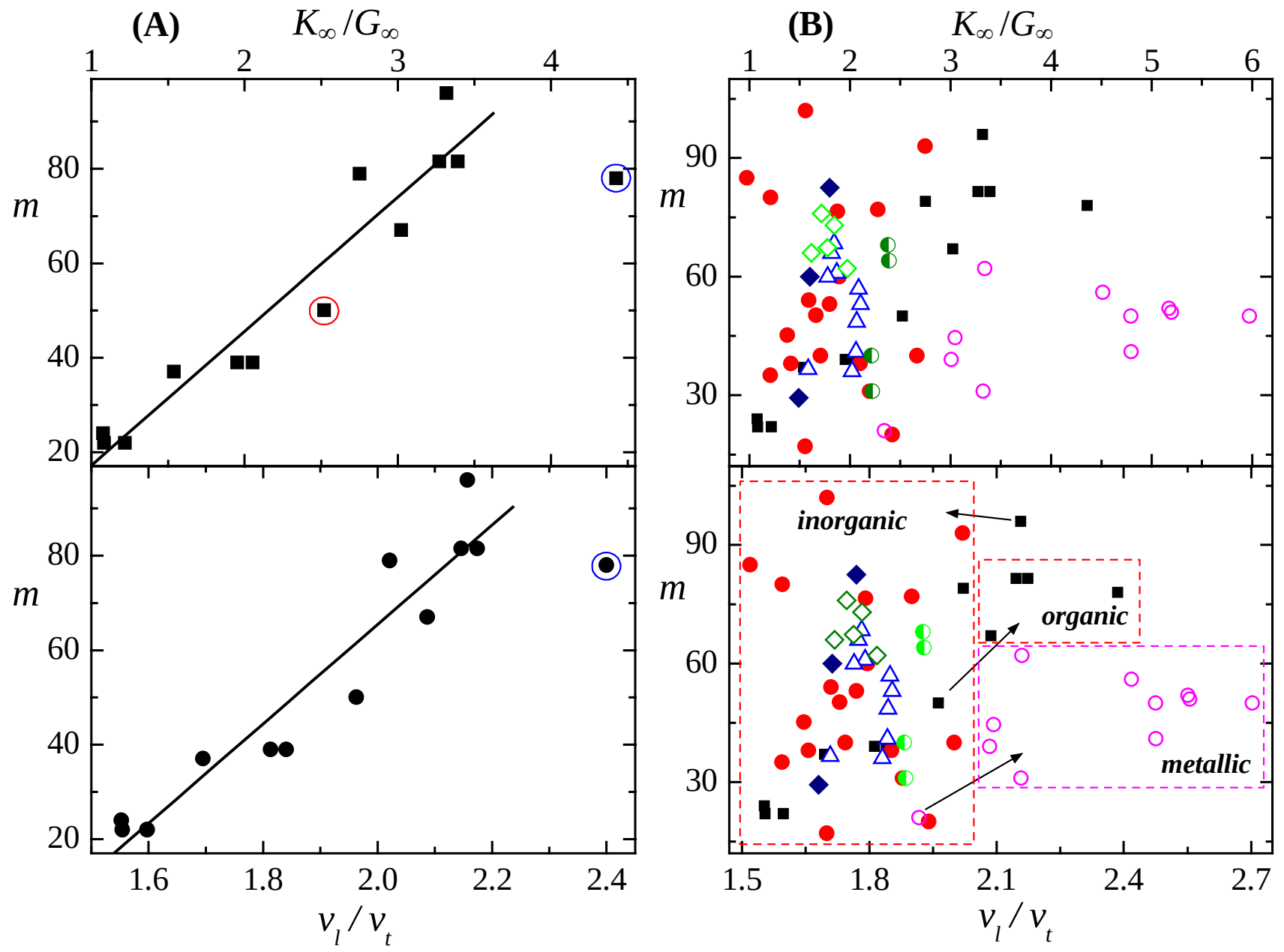

**Figure 1**